\documentclass[aps,pra,twocolumn,superscriptaddress,reprint]{revtex4-2}

\usepackage[babel,threshold=2]{csquotes}
\usepackage[utf8]{inputenc}
\usepackage[T1]{fontenc}
\usepackage{amsmath}
\usepackage{amssymb}
\usepackage{amsthm}

\usepackage{color}
\usepackage{comment}
\usepackage{bbold}
\usepackage{mathtools}
\usepackage{graphicx}
\usepackage{hyperref}
\usepackage{epsfig}
\usepackage[normalem]{ulem}
\usepackage{amsfonts}
\usepackage{amssymb}
\usepackage{enumerate}
\usepackage{amsmath}
\usepackage{amsthm}
\usepackage{graphicx}
\usepackage[dvipsnames]{xcolor}
\usepackage[english]{babel}
\usepackage{comment}
\usepackage{tensor}
\usepackage{hyperref}
\usepackage{multirow}
\usepackage{mathtools} 
\usepackage{graphicx}    
\usepackage{subcaption}  
\usepackage{braket} 
\usepackage{float}
\usepackage{tikz}
\usepackage{soul}
\usetikzlibrary{arrows.meta,decorations.markings}
\usetikzlibrary{arrows,decorations.pathmorphing,positioning}

\begin{document}

\title{Quantitative comparison of quantum pseudo-telepathy games and Bell inequalities}

\author{G\'abor Homa}
\email{homa.gabor@wigner.hun-ren.hu}
\address{ HUN-REN Wigner Research Centre for Physics, Konkoly-Thege M. \'ut 29-33, H-1121 Budapest, Hungary}

\author{Andr\'as Bodor} 
\address{ HUN-REN Wigner Research Centre for Physics, Konkoly-Thege M. \'ut 29-33, H-1121 Budapest, Hungary}
\address{Institute of Mathematics and Informatics, Faculty of Science,
University of P\'ecs, H-7624 P\'ecs, Ifj\'us\'ag \'utja 6, Hungary}

\author{J\'ozsef Zsolt Bern\'ad}
\email{j.bernad@fz-juelich.de}
\address{ HUN-REN Wigner Research Centre for Physics, Konkoly-Thege M. \'ut 29-33, H-1121 Budapest, Hungary}

\begin{abstract}
Quantum pseudo-telepathy games, such as the Mermin–Peres magic square and the doily game, theoretically allow players to win with unit probability when using entangled quantum strategies. We quantitatively characterize the quantum advantage in these games and compare it with violations of two Bell inequalities: the Clauser-Horne–Shimony–Holt and the Collins–Gisin inequalities. The analysis is restricted to two families of two-qubit states: modified Werner states and Bell-diagonal states. For each case, we identify and quantify the regions of quantum state space that exhibit either a quantum advantage or a Bell inequality violation, relative to the set of all entangled states. Within these families, the doily game encompasses a larger fraction of entangled states than the Mermin–Peres magic square game, although both cover significantly smaller regions of entangled states compared to those where Bell inequalities are violated. Although both approaches are fundamentally linked to quantum contextuality, our analysis of the examined two-qubit state families suggests that Bell inequalities are more effective at revealing entanglement, even if pseudo-telepathy games offer a more intuitive and conceptually appealing perspective.

\end{abstract}

\maketitle
\section{Introduction}
The rapid evolution of quantum information science has placed entanglement at the heart of numerous applications \cite{Leuchs,Arute,Zhong,Marciniak2022}. It is interesting that entanglement, regarded as the most distinctive departure from classical physics in quantum theory, was employed by Einstein, Podolsky, and Rosen (EPR) in their attempt to assign definite values to physical properties before any measurement was made \cite{EPR}. Quantum mechanics challenges this view and shows that the outcome of a measurement can depend on which other properties are measured at the same time. This concept, known as quantum contextuality, reveals that quantum behavior cannot always be explained using classical ideas. For a recent and comprehensive overview of this topic, see \cite{Guehne}. A well known approach by Bohm presented a classically motivated hidden variable (HV) model which, due to its nonlocal structure, is contextual \cite{Bohm1, Bohm2}. Later, Kochen and Specker \cite{Specker, KS}, as well as Bell \cite{Bell1, Bell2}, showed that it is impossible to reproduce quantum mechanics using noncontextual HV models. It is important to note that Kochen and Specker showed the hidden variable problem admits a positive solution in a two-dimensional complex Hilbert space \cite{KS}, a result motivated by Gleason's Theorem \cite{Gleason}.  

A different approach by Greenberger, Horne, and Zeilinger shed a new light on the relation between the original EPR argument and Bell's approach \cite{GHZ}. Motivated by this result, Mermin and Peres introduced the so-called Mermin–Peres magic square \cite{Peres, Mermin1, Mermin2}. This construction consists of nine measurements, composed of single-qubit Pauli matrices of two two-level systems, arranged in a square, which demonstrates the impossibility of noncontextual HV models. The magic square can be reformulated as a game \cite{Aravind1, Aravind2}, giving rise to a two-player pseudo-telepathy scenario. Quantum pseudo-telepathy refers to a class of quantum games in which players, by using quantum strategies, can win with certainty, i.e., with probability one \cite{Brassard2005}.
Moreover, the Mermin-Peres magic square game (MPMG) has been experimentally realized using a photonic setup \cite{Xu}. The Mermin–Peres magic square admits further generalizations, such as the Mermin pentagram \cite{Mermin2}, which provides a three-qubit, observable-based proof of quantum contextuality. These constructions have been extensively studied from the perspective of finite geometry. In Ref. \cite{Saniga1}, it was demonstrated that Mermin’s configurations are the most economical in terms of the number of observables. Furthermore, Ref. \cite{Saniga2} showed that the ten possible Mermin–Peres magic squares correspond to hyperplanes in a point–line geometry known as the doily. This framework can also be interpreted as a quantum pseudo-telepathy game, i.e., doily game (DG) \cite{Kelleher, Kelleher_2024}. For further connections between contextuality and geometry, see Ref. \cite{Boutray_2022}.

In this paper, we focus on the quantitative characterization of the two aforementioned approaches: Bell inequalities, which constrain the possible correlations between spatially separated particles, and quantum pseudo-telepathy games. Assuming the validity of quantum mechanics, we examine how many quantum states are capable of violating Bell inequalities or outperforming classical strategies in these games. The quantitative evaluation refers to the  measure ratios inside the convex body of all quantum states. As a starting point, we revisit earlier work by one of the authors \cite{Sauer_2021}, where the Clauser–Horne–Shimony–Holt (CHSH) \cite{ClauserHorne1974} and Collins–Gisin (CG) \cite{CG} inequalities were investigated. Here, we compare these inequalities with the MPMG and DG. Furthermore, we examine two families of two-qubit states: a modified Werner state \cite{Werner} and the Bell-diagonal states. Since these states are characterized by one or three parameters, our results lend themselves to geometric visualization. It is worth noting that the behavior of the MPMG under noise has been previously studied in Ref. \cite{Fialik}. However, in this work, our focus is on the typicality of both mixed and pure entangled states that are sufficient to outperform classical strategies in the games. We do not explore the origins of not-perfectly entangled states, such as whether the states have been affected by noise or other imperfections during their preparation or processing. In this context, we demonstrate that a larger number of states can achieve success in the DG compared to the MPMG. Moreover, when compared with the selected Bell inequalities, we find that the set of states capable of violating these inequalities is generally larger than the set of states that can outperform classical strategies in the corresponding quantum games.

The structure of the paper is as follows. In Sec. \ref{II}, we introduce the two quantum pseudo-telepathy games under consideration, along with the corresponding classical strategies and their winning probabilities. Sec. \ref{III} presents the theoretical and mathematical background necessary for our analysis. The results are discussed in Sec. \ref{IV}. Finally, Sec. \ref{V} summarizes our conclusions. Additional technical details are provided in the Appendices.

\section{Technical Description of the Games}
\label{II}

There are two equivalent frameworks for addressing the impossibility of noncontextual HV models: Bell inequalities and nonlocal games. In the first approach, one derives linear inequalities that any measurement statistics must satisfy if they are to be explained by a noncontextual HV model. These inequalities can be represented as facets of classical correlation polytopes \cite{Pitowsky}. A more intuitive and conceptually straightforward approach is through nonlocal games \cite{Brassard2005}. A two-player nonlocal game consists of two sets of inputs/questions and outputs, a predicate called the promise, and a winning condition, which is a relation between the inputs and outputs. The two cooperating players, typically named Alice and Bob, must satisfy this relation whenever the promise holds. In a deterministic classical approach, the behavior of Alice and Bob is fully described by two fixed functions mapping their respective inputs to outputs. The effectiveness of such a strategy is measured by the fraction of valid input/question pairs for which it yields a correct response, assuming the inputs/questions are given/asked uniformly at random among all legitimate ones. If the players are permitted to use private randomness and to access shared random variables, they can employ probabilistic methods. In this scenario, the focus shifts to the likelihood that their strategy produces a correct answer for specific inputs. As demonstrated in Ref. \cite{Brassard2005}, the best possible success probability achieved through probabilistic strategies is always less than or equal to the highest success rate attained by some deterministic strategy. In contrast, employing quantum resources allows for greater chances of winning, as demonstrated by the Mermin-Peres magic square game (MPMG).

Before we explain the games studied in this paper, we briefly describe how the CHSH inequality, viewed as a game, can provide better winning probabilities. In this scenario, two parties, Alice and Bob, each receive Boolean inputs $x$ and $y$, respectively, chosen uniformly at random, and produce Boolean outputs $a$ and $b$. The goal is to satisfy the winning condition $a\oplus b = x\land y$, where $\oplus$ and $\land$ denote the XOR and AND operations, respectively. Without entanglement, the best success rate is $0.75$. Sharing entangled states raises this to $(2+\sqrt{2})/4 \approx 0.85$ \cite{Brunner}. Classical strategies rely on predetermined answers for each input pair; there are $16$ such deterministic strategies in total. As mentioned above, using shared randomness to combine these strategies does not increase the maximum winning probability beyond that of the best deterministic strategy.

In Refs. \cite{Aravind1, Aravind2}, Aravind introduced a nonlocal quantum game inspired by the Mermin–Peres magic square, offering a demonstration of quantum nonlocality without relying on inequalities \cite{Hardy, Cabello}. In this game, there are three possible questions and eight possible answers. The game is based on a
$3 \times 3$ matrix filled with entries from the set $\{1,-1\}$. Alice is asked to provide the values in a selected row, while Bob is asked to provide the values in a selected column. To win the game, Alice and Bob must ensure that their answers agree on the single cell where the chosen row and column intersect. Additionally, the product of Alice’s three returned values must equal $1$, and the product of Bob’s must equal $-1$. However, since there is no such matrix that can satisfy all these conditions simultaneously \cite{Aravind1}, it is impossible for classical players to always win. Therefore, in the classical setting, disagreement between Alice’s and Bob’s answers is unavoidable in some cases. In an optimal classical strategy, Alice and Bob can each use matrices that satisfy their respective parity constraints, differing in only one of the nine entries. When the questions are chosen uniformly at random, there is a one in nine chance that they are asked about the conflicting entry. Consequently, the maximum success rate in the classical case is $8/9$.

\begin{table}[ht!]
    \centering
    \begin{tabular}{|c|c|c|}
    \hline
     $I \otimes Z $  & $ Z \otimes I $  & $ Z \otimes Z $ \\ \hline
    $ X \otimes I $  & $ I \otimes X $  & $ X \otimes X $\\ \hline
     $ - X \otimes Z $ & $- Z \otimes X $ & $ Y \otimes Y $ \\
   \hline 
    \end{tabular}
    \caption{An example of the Mermin–Peres magic square game (MPMG). Each entry corresponds to a two-qubit Pauli observable. Observables within the same row or column commute, while those outside of these alignments anti-commute.}
    \label{peres_mermin_game}
\end{table}

In contrast to the CHSH game, the MPMG is a quantum pseudo-telepathy game, meaning there exists a quantum strategy that guarantees a win with certainty. In this setup, Alice and Bob share two pairs of maximally entangled qubits, each one in the state $(\ket{00}+\ket{11})/\sqrt{2}$. Their responses are determined by measurement outcomes corresponding to carefully chosen combinations of Pauli observables, that is, the Pauli matrices $X$, $Y$, and $Z$ together with the identity matrix $I$. The specific measurement settings are listed in Table~\ref{peres_mermin_game}. For example, to determine the value of the matrix entry in position $(3,3)$, Alice (or Bob) ought to measure $Y\otimes Y$,
where the tensor product is with respect to the two
qubits on Alice's (or Bob's) side. While it might seem intuitive to simply measure each qubit in the $Y$ basis and multiply the outcomes, this approach is incorrect. Alice is required to provide values for all entries in the last row. Since the three observables in each row and column commute, they can, in principle, be measured simultaneously. However, their common eigenvectors are entangled across the two qubits, which they have at hand, meaning the measurement cannot be performed independently on each qubit. Instead, it must involve a joint measurement that acts nontrivially on both qubits. Therefore, the measurement for the last row will be performed by Alice in the following orthonormal basis:
\begin{eqnarray}
\phi_{1,1,1} &=&\frac{1}{2} \begin{pmatrix}1\\-1\\-1\\-1\end{pmatrix}, \quad
\phi_{1,-1,-1} = \frac{1}{2} \begin{pmatrix}1\\1\\-1\\1\end{pmatrix} \nonumber \\
\phi_{-1,-1,1} &=& \frac{1}{2} \begin{pmatrix}1\\1\\1\\-1\end{pmatrix}, \quad 
\phi_{-1,1,-1} = \frac{1}{2} \begin{pmatrix}1\\-1\\1\\1\end{pmatrix},
\end{eqnarray}
where the lower indices of $\phi$ indicate the possible entry values of the last row, ordered from left to right. Naturally, performing measurements in this basis is significantly more demanding in terms of hardware than measuring individual qubits. However, the outcomes automatically satisfy the constraint that their product equals $+1$. It is also straightforward to verify, that if Alice and Bob measures the same operator (out of these nine) on their share of the qubits, they always obtain identical outcomes. Therefore, the winning condition is always satisfied.\\

\begin{figure}[ht]
\centering
\begin{tikzpicture}[scale=0.96, every node/.style={font=\small}]
\coordinate (P1) at (0,4);
\coordinate (P2) at (3.80423, 1.23607);
\coordinate (P3) at (2.35114, -3.23607);
\coordinate (P4) at (-2.35114, -3.23607);
\coordinate (P5) at (-3.80423, 1.23607);

\coordinate (P6)  at (1.90212,2.61804);
\coordinate (P7)  at (3.07769,-1); 
\coordinate (P8)  at (0,-3.23607); 
\coordinate (P9)  at (-3.07769,-1); 
\coordinate (P10) at (-1.90212,2.61804); 

\coordinate (P11) at 0.75*(0,-2);
\coordinate (P12) at 0.75*(-1.90212,-0.61804); 
\coordinate (P13) at 0.75*(-1.17557,1.61804); 
\coordinate (P14) at 0.75*(1.17557,1.61804);
\coordinate (P15) at 0.75*(1.90212,-0.61804);

\foreach \a/\b in {P1/P6, P6/P2, P2/P7, P7/P3, P4/P9, P9/P5, P5/P10, P10/P1,
                   P1/P8, P6/P4, P7/P5, P3/P10, P2/P9} {
  \draw[thick,shorten >=1.5pt,shorten <=1.5pt] (\a) -- (\b);
}


\foreach \a/\b/\c in {P11/P9/P13,  P14/P7/P11, P15/P8/P12} {
  \draw[thick,shorten >=1.5pt,shorten <=1.5pt] plot [smooth, tension=0.25] coordinates {(\a)  (\b)  (\c)};
}

\foreach \a/\b in {P3/P4} {
  \draw[red,thick,shorten >=1.5pt,shorten <=1.5pt] (\a) -- (\b);
}


\foreach \a/\b/\c in {P12/P10/P14,  P13/P6/P15} {
  \draw[red,thick,shorten >=1.5pt,shorten <=1.5pt] plot [smooth,tension=0.25] coordinates {(\a)  (\b)  (\c)};
}

\tikzset{lbl/.style={fill=white, inner sep=1.2pt, outer sep=0pt}}

\node[lbl] at (P1)  {$I \otimes Y$};
\node[lbl] at (P2)  {$Z \otimes I$};
\node[lbl] at (P3)  {$Z \otimes Z$};
\node[lbl] at (P4)  {$X \otimes X$};
\node[lbl] at (P5)  {$X \otimes I$};

\node[lbl] at (P6)  {$Z \otimes Y$};
\node[lbl] at (P7)  {$I \otimes Z$};
\node[lbl] at (P8)  {$Y \otimes Y$};
\node[lbl] at (P9)  {$I \otimes X$};
\node[lbl] at (P10) {$X \otimes Y$};

\node[lbl] at (P11) {$Y \otimes I$};
\node[lbl] at (P12) {$Z \otimes X$};
\node[lbl] at (P13) {$Y \otimes X$};
\node[lbl] at (P14) {$Y \otimes Z$};
\node[lbl] at (P15) {$X \otimes Z$};

\end{tikzpicture}
\caption{The structure of the doily game (DG). Possible questions correspond to two-qubit Pauli observables connected by solid lines and curves. Observables lying on the same line or curve commute. Black lines and curves represent positive parity questions, while red lines indicate negative parity questions.}
\label{fig:doily_pauli}
\end{figure}
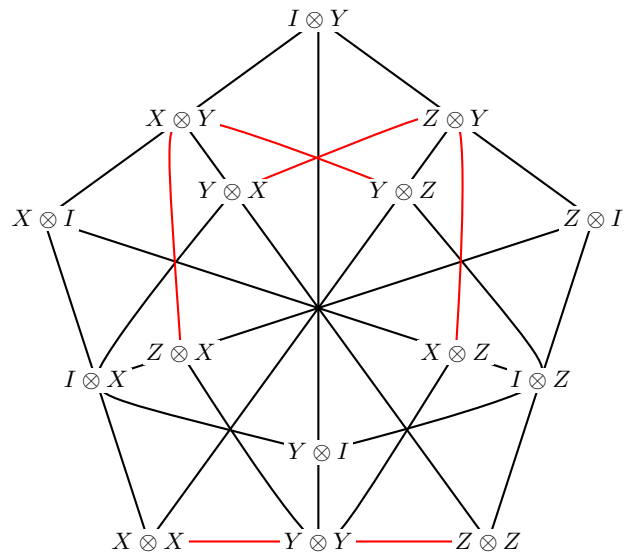

The doily game (DG) is a generalization of the MPMG. While the magic square game uses $9$ specific combinations of Pauli matrices, there are actually $15$ nontrivial combinations--so why not use all of them? In this game, Alice and Bob are each given a maximal set of commuting operators. There are 15 such sets, each containing $3$ operators. Interestingly, the product of the operators in each set is either the identity operator (in 12 cases) or minus the identity operator (in 3 cases), as shown in Fig. \ref{fig:doily_pauli}. The questions are selected so that Alice’s and Bob’s sets share exactly one operator. Each player must respond with values for all the operators in their set. For the shared operator, they must provide the same result. Additionally, the product of their three answers must equal either $+1$ or $-1$, depending on the sign of the product of the corresponding operators. An optimal classical strategy is to return all ones if the parity of the selected operators is $+1$, and to randomly select one operator to assign the value $-1$ if the parity is $-1$. There are $15 \times 6$ possible questions. Among these, in $9 \times 4$ cases, one of the players receives a question with negative parity. In these cases, the players lose with probability $1/3$. Therefore, the maximum success rate in the classical case is $13/15= 0.8\overline{6}$, as shown in \cite{Kelleher}. The latter reference discusses another game based on the geometric structure of the doily; however, its maximum success rate is $14/15$, which becomes a disadvantage when the game is played using entangled states.

The quantum winning strategy closely resembles that of the MPMG. Alice and Bob share two pairs of maximally entangled qubits, i.e., the $(\ket{00}+\ket{11})/\sqrt{2}$ state, and perform measurements of the specified operators on their respective halves. One key difference is that while measuring $X$ or $Z$ yields correlated results for Alice and Bob, measuring $Y$ produces anticorrelated outcomes. Therefore, for operators containing an odd number of $Y$ terms, one player applies the additive inverse to their measurement outcome. (This adjustment is not necessary in the MPMG.) Following this procedure ensures that all the game’s constraints are satisfied. It is worth noting that these types of games can be further generalized in various ways, expanding both their theoretical foundations and practical applications. For a deeper exploration of these extensions, see, for example, Refs. \cite{Kelleher_2024, arkhipov, Adamson, Trandafir_2022}, which discuss broader classes of games that push beyond the frameworks presented here.

\section{Theoretical Background}
\label{III}

In this section, we introduce the mathematical framework used to analyze both the quantum games under study and the Clauser–Horne–Shimony–Holt (CHSH) and Collins–Gisin (CG) inequalities. Our analysis focuses on determining the fraction of entangled states that either provide a quantum advantage in the games or violate the Bell inequalities.

\subsection{Theoretical framework in quantum physics} 
\label{sec:theoretical_framework}

As discussed in the previous section, if Alice and Bob share two entangled states of the form $(\ket{00}+\ket{11})/\sqrt{2}$, they can win both the MPMG and the DG with certainty. In the following, we examine specific families of two-qubit states. The first is a modified Werner state, characterized by a single real parameter that describes its fidelity with respect to the
$(\ket{00}+\ket{11})/\sqrt{2}$ state, rather than the more commonly used $(\ket{01}-\ket{10})/\sqrt{2}$ state. We then extend our analysis to Bell-diagonal states, which are described by three independent real parameters. Our central question is: under what conditions can these states outperform the classical maximum success winning rate of the games?

The Bell states are defined as follows
\begin{equation}
\ket{\Phi^{\pm}}=\frac{\ket{00} \pm \ket{11}  }{\sqrt{2}}, \quad \ket{\Psi^{\pm}}=\frac{\ket{01} \pm \ket{10}  }{\sqrt{2}}.
\end{equation}
Let the modified Werner state with non-negative real parameter $F$ be:
\begin{eqnarray}
\rho(F)&:=&F \ket{\Phi^+}\bra{\Phi^+}+\frac{1-F}{3} \Big( \ket{\Phi^-}\bra{\Phi^-}  \nonumber \\
 &+&\ket{\Psi^+}\bra{\Psi^+}+ \ket{\Psi^-}\bra{\Psi^-} \Big),
 \label{eq:Werner}
\end{eqnarray}
where $F$ denotes the fidelity with respect to the
$\ket{\Phi^{+}}$ state. Bell-diagonal states are characterized by three independent real-valued parameters and take the form:
\begin{eqnarray}
\label{eq:Belldiagonal}
\rho(a,b,c)&:=&a \ket{\Phi^+}\bra{\Phi^+} +b \ket{\Psi^+}\bra{\Psi^+}  \\ 
&+& c \ket{\Phi^-}\bra{\Phi^-}+ (1-a-b-c) \ket{\Psi^-}\bra{\Psi^-}, \nonumber 
\end{eqnarray}
where $0\leqslant a,b,c\leqslant 1$ and $a+b+c \leqslant 1$. Alice and Bob are given two copies of qubit pairs, either $\rho=\rho(F)\otimes \rho(F)$ or $\rho=\rho(a,b,c) \otimes \rho(a,b,c)$. According to the Peres–Horodecki criterion, in the special case of two-qubit systems, all PPT (positive partial transpose) quantum states are separable \cite{Peressep, Horodeckisep}. In other words, bound entanglement, that is, entangled states with positive partial transpose, does not occur in these cases. For modified Werner states, $\rho(F)$ is entangled when $F>0.5$.
The set of Bell-diagonal states forms a tetrahedron, within which the separable states occupy an octahedral region. See, for example, Ref. \cite{Sauer_2021} and the references therein. It is worth noting that the modified Werner states form a subset of the Bell-diagonal states. Within the tetrahedral representation, this subset corresponds to a line segment that starts at the vertex associated with $\ket{\Phi^+}$, passes through the center of the tetrahedron, and ends at the opposite triangular face of that vertex.

The measurement operators used in the MPMG protocol are given by: 
\begin{eqnarray}
\mathcal{O}\in& \mathcal{M}_{\text{MPMG}} = \{ I\otimes Z, Z\otimes I, Z \otimes Z,  
X \otimes I, I \otimes X, \nonumber \\ 
&X \otimes X, 
-X \otimes Z, -Z \otimes X, Y \otimes Y \},
\end{eqnarray}
while for the DG protocol, the set of measurement operators is:
\begin{eqnarray}
\mathcal{O}\in& \mathcal{M}_{\text{DG}} =  \{ I\otimes Y, Z\otimes I, Z \otimes Z,  
X \otimes X, X \otimes I, \nonumber \\ 
&Y \otimes Y, I \otimes X, X \otimes Y, Z \otimes Y, I \otimes Z \nonumber \\
&Y \otimes I, Z \otimes X, Y \otimes X, Y  \otimes Z, X \otimes Z \}, 
\end{eqnarray}
where $X,Y,Z$ are the Pauli matrices and $I$ is the $2\times2$ identity matrix.

Now, consider that Alice performs measurements using the local observables $\mathcal{O}^{A_1}$, $\mathcal{O}^{A_2}$, and $\mathcal{O}^{A_3}$, obtaining outputs $a_1$, $a_2$, and $a_3$, respectively. Simultaneously, Bob measures
$\mathcal{O}^{B_1}$, $\mathcal{O}^{B_2}$, and $\mathcal{O}^{B_3}$ and obtains outputs $b_1$, $b_2$, and $b_3$. As discussed in Sec. \ref{II}, the measurement outcomes can take values in in $\{1,-1\}$, and the observables belong to the set $\mathcal{M}_X$, where $X \in \{\text{MPMG}, \text{DG}\}$. Moreover, the measurement operators satisfy the commutation relation:
\begin{equation}
\left[ \mathcal{O}^{\Gamma_{i}},\mathcal{O}^{\Gamma_{j}}\right]=0, \quad \Gamma\in \{A,B\} \,\, \text{and} \, \, i,j=1,2,3.
\end{equation}
Therefore, the observables $\mathcal{O}^{A_1}$, $\mathcal{O}^{A_2}$, and $\mathcal{O}^{A_3}$ are simultaneously diagonalizable, and their spectra consist solely of the eigenvalues $\pm 1$. For a given set of outputs $a_1$, $a_2$, and $a_3$, there exists a corresponding projector $\Pi_{\mathcal{O}^{A_1},\mathcal{O}^{A_2},\mathcal{O}^{A_3}}^{a_1,a_2,a_3}$ that projects onto the joint eigenspace associated with these results. The same reasoning applies analogously on Bob’s side. Since the entire process depends on the choice of observables for both Alice and Bob, as well as on the corresponding outcomes, we denote this event as:
\begin{eqnarray}
 \mathbf{W}= (a_1,a_2,a_3,b_1,b_2,b_3,\mathcal{O}^{A_1},\mathcal{O}^{A_2},\mathcal{O}^{A_3},\mathcal{O}^{B_1},\mathcal{O}^{B_2},
 \mathcal{O}^{B_3}) \nonumber \\ 
\end{eqnarray}
with probability 
\begin{eqnarray}
P(\mathbf{W})=\mathrm{Tr}\left\{  \left(\Pi_{\mathcal{O}^{A_1},\mathcal{O}^{A_2},\mathcal{O}^{A_3}}^{a_1,a_2,a_3} \otimes \Pi_{\mathcal{O}^{B_1},\mathcal{O}^{B_2},\mathcal{O}^{B_3}}^{b_1,b_2,b_3} \right)  \rho\right\}. \nonumber \\
\end{eqnarray}    
The probability of winning the game for a fixed question is
\begin{equation}
w(\mathcal{O}^{A_1},\mathcal{O}^{A_2},\mathcal{O}^{A_3},\mathcal{O}^{B_1},\mathcal{O}^{B_2},\mathcal{O}^{B_3})=\sum_{a_i,b_j \text{admissible}} P(\mathbf{W}).
\end{equation}

In the case of the MPMG, a combination of outputs $a_i$ and $b_j$ is considered admissible if the product of Alice’s outputs satisfies $a_1a_2a_3=1$ , the product of Bob’s outputs satisfies $b_1b_2b_3=-1$, and the measurement outcomes for any shared observable between Alice and Bob are identical. For the DG, a combination of $a_i$ and $b_j$ is admissible if the products of the outcomes satisfy the conditions specified in Sec. \ref{II}. These conditions depend on the positions of the observables in Fig. \ref{fig:doily_pauli}, and, as in the MPMG case, the measurement outcomes for any common observable between Alice and Bob must match.

The average probability of winning the game is given by
\begin{eqnarray}
    \overline{w}=&\sum p(\mathcal{O}^{A_1},\mathcal{O}^{A_2},\mathcal{O}^{A_3},\mathcal{O}^{B_1},\mathcal{O}^{B_2},\mathcal{O}^{B_3})\nonumber\\
&w(\mathcal{O}^{A_1},\mathcal{O}^{A_2},\mathcal{O}^{A_3},\mathcal{O}^{B_1},\mathcal{O}^{B_2},\mathcal{O}^{B_3})
\end{eqnarray}
where $p(\dots)$ denotes the probability of a given question, which in our case corresponds to a uniform distribution over the allowed operator selections. In the MPMG, Alice’s operators are chosen from a row of Table \ref{peres_mermin_game}, and Bob’s operator from a column, resulting in $p(\dots)=\frac{1}{9}$. In the DG, Alice's and Bob's operator sets must share exactly one common element, and all operators within each set must commute, see Fig. \ref{fig:doily_pauli}. Under these constraints, the probability is $p(\dots)=\frac{1}{90}$.

Finally, let 
\begin{equation}
\mathcal{R} \subset \mathcal{E} \subset \mathcal{S}
\end{equation}
denote a sequence of nested sets of quantum states, where:
\begin{itemize}
    \item $\mathcal{S}$ is the convex set of all quantum states under consideration.
    \item $\mathcal{E} \subset \mathcal{S}$ denotes the subset of entangled states.
    \item $\mathcal{R} \subset \mathcal{E}$ is the subset of entangled states for which quantum games yield a higher winning probability than any classical strategy. This notation is also used for states that violate Bell inequalities.
\end{itemize}

Based on Eqs. \eqref{eq:Werner} and \eqref{eq:Belldiagonal}, the set $S$ corresponds either to an interval in one-dimensional Euclidean space or to a tetrahedron in three-dimensional Euclidean space. Consequently, $\mathcal{S}$ is Lebesgue measurable, and its measure can be computed using the 
$n$-dimensional Lebesgue measure (here $n=1$ or $3$) \cite{stein2005real}:
\begin{equation}
V(\mathcal{S}) := m_n(\mathcal{S}).
\end{equation}
In order to treat the sets $\mathcal{R}$ and $\mathcal{E}$, in a unified manner, we consider:
\begin{equation}
\mathcal{A} = \{ \mathbf{x} \in \mathbb{R}^n \mid \varphi(\mathbf{x}) \},
\end{equation}
where $\varphi(\mathbf{x})$ is a logical statement or condition on $\mathbf{x}$, which returns a Boolean value. In our case, the condition represents whether a given state is entangled, achieves a higher winning probability than any classical strategy in a quantum game, or violates a Bell inequality. 

The measure of the set $\mathcal{A}$ can be equivalently represented using the indicator function and the  $n$-dimensional Lebesgue measure as follows:
\begin{equation}
V(\mathcal{A})  = m_n(\mathcal{A}) =  \int_{\mathbb{R}^n} \mathbf{1}_\mathcal{A} (\mathbf{x}) \, d^n\mathbf{x},
\end{equation}
where $\mathbf{1}_\mathcal{A} (\mathbf{x})$ denotes the indicator function associated with the set $\mathcal{A}$, defined by:
\begin{equation}
\mathbf{1}_\mathcal{A} (\mathbf{x}) = 
\begin{cases}
1, & \text{if } \varphi(\mathbf{x}) \text{ holds}, \\
0, & \text{otherwise}.
\end{cases}
\end{equation}
Now, $\mathcal{A}$ denotes one of the sets $\mathcal{R}$, $\mathcal{E}$ or $\mathcal{S}$, where $\mathcal{R} \subset \mathcal{E} \subset \mathcal{S}$. We are interested in analyzing the following two ratios:
\begin{equation}
\label{eq:Pr}
P(\mathcal{R}) = \frac{V(\mathcal{R})}{V(\mathcal{S})},
\end{equation}
and
\begin{equation}
\label{eq:PrQ}
P_Q(\mathcal{R}) = \frac{V(\mathcal{R})}{V(\mathcal{E})}.
\end{equation}
One of the primary objectives of this study is to compute and compare these ratios for the two quantum games and the two Bell inequalities, which are discussed in the subsequent section.

\subsection{Bell inequalities}

The purpose of this section is to review and summarize established results concerning CHSH and CG inequalities. Additionally, we present the corresponding condition for Bell-diagonal states under which no experimental configuration can lead to a violation of either the CHSH or the CG inequality.

A general two-qubit density matrix can be expressed in the form
\begin{eqnarray}
&& \rho_{\text{AB}} = \frac{I^{(A)} \otimes I^{(B)}}{4} +
  \frac{1}{2}\sum_{N=X,Y,Z}\tau^{(A)}_N N^{(A)} \otimes I^{(B)}  \nonumber \\ &&+\frac{1}{2}\sum_{N=X,Y,Z}\tau^{(B)}_N I^{(A)} \otimes N^{(B)}  \label{eq:genrho}\\
  &&+
 \frac{1}{2}\sum_{N,M=X,Y,Z}\nu_{NM} N^{(A)} \otimes M^{(B)}, \nonumber 
\end{eqnarray}
where $\tau^{(A)}_N$, $\tau^{(B)}_N$, and $\nu_{NM}$ are real parameters for all $N,M \in \{X,Y,Z\}$. This representation ensures that the matrix $\rho_{\text{AB}}$ is self-adjoint and has unit trace. Additional constraints on these real parameters are required to guarantee the positive semidefiniteness of $\rho_{\text{AB}}$ \cite{Sauer_2021}. The real parameters $\nu_{NM}$ form a $3 \times 3$ matrix, which we denote by $C_\nu$.

The CHSH inequality~\cite{ClauserHorne1974} pertains to bipartite correlation experiments involving two spatially separated parties, $A$ and $B$, each of whom can choose between two measurement settings, with binary outcomes $\pm 1$. This scenario involves four observables: $A_1$, $A_2$ for party $A$, and $B_1$, $B_2$ for party $B$. In terms of these observables, the CHSH inequality is expressed as follows:
\begin{eqnarray}
-2 \leqslant \mathbb{E}(A_1 B_1) + \mathbb{E}(A_1 B_2) + \mathbb{E}(A_2 B_1) - \mathbb{E}(A_2 B_2) \leqslant 2, \nonumber \\
\label{eq:chshineq}
\end{eqnarray}
with $\mathbb{E}(.)$ denoting the expectation value. Based on the results presented in Ref. \cite{horodecki_family_95}, the CHSH inequality can equivalently be expressed as
\begin{equation}
\lambda_1 + \lambda_2 \leqslant \frac{1}{4},
\label{eq:horodeckicond}
\end{equation}
where $\lambda_1$ and $\lambda_2$ are the two largest eigenvalues of the matrix $C^\dag_\nu C_\nu$. Accordingly, for any quantum state $\rho_{\text{AB}}$ satisfying this condition, no choice of measurement settings for the four observables can lead to a violation of the CHSH inequality. 

For Bell-diagonal states of the form given in \ref{eq:Belldiagonal}, we have $\tau^{(A)}_N=\tau^{(B)}_N=0$ for all $N=X,Y,Z$. In this case, the correlation matrix $C_\nu$ takes the diagonal form: 
\begin{equation}
\label{eq:Cn}
C_\nu= \begin{pmatrix}
a+b-\frac{1}{2} & 0 & 0 \\
0 & b+c-\frac{1}{2}& 0 \\
0 & 0 & a+c-\frac{1}{2}
\end{pmatrix}.
\end{equation}
Substituting this into the condition from \eqref{eq:horodeckicond} leads to the following three inequalities:
\begin{eqnarray}
\left(a+b-\frac{1}{2}\right)^2+\left( b+c-\frac{1}{2}\right)^2 \leqslant \frac{1}{4}, \nonumber \\
\left(a+b-\frac{1}{2}\right)^2+\left(a+c-\frac{1}{2}\right)^2 \leqslant \frac{1}{4}, \label{eq:abcCHSH}  \\
\left( b+c-\frac{1}{2}\right)^2+\left(a+c-\frac{1}{2}\right)^2 \leqslant \frac{1}{4}. \nonumber
\end{eqnarray}
If we temporarily omit the fact that the three parameters are subject to conditions ensuring the positive semidefiniteness and unit trace of $\rho(a,b,c)$ in Eq.~\eqref{eq:Belldiagonal}, then the three inequalities collectively define a geometric body known as the Steinmetz solid (or tricylinder), which corresponds to the intersection of three identical cylinders oriented along mutually orthogonal axes \cite{Sauer_2021}. However, if the conditions ensuring that $\rho(a,b,c)$ is a valid quantum state are imposed, one must consider the intersection of the tetrahedron (see Sec. \ref{III}) and the Steinmetz solid, which has parts lying outside the tetrahedron.

When three distinct measurement settings are available at both sites $A$ and $B$, each yielding binary outcomes, Collins and Gisin \cite{CG} derived a new class of Bell-type inequalities that is based on the work of Pitowsky and Svozil \cite{Svozil}. In addition to modified forms of the CHSH inequality, their formulation includes a fundamentally new CG inequality, given by:
\begin{eqnarray} 
 &&0\leqslant 4 +  \mathbb{E}(A_1) + \mathbb{E}(A_2) + \mathbb{E}(B_1)+\mathbb{E}(B_2)+  \mathbb{E}(A_1B_1) \nonumber \\
 &&+ \mathbb{E}(A_1B_2)+\mathbb{E}(A_2B_1)+ \mathbb{E}(A_2B_2)+ \mathbb{E}(A_1B_3)+ \mathbb{E}(A_3B_1) \nonumber \\
&&-\mathbb{E}(A_2B_3)-\mathbb{E}(A_3B_2). \label{eq:CG}
\end{eqnarray}
In the case of Bell-diagonal quantum states of the form given in Eq. \eqref{eq:Belldiagonal}, this inequality can be equivalently reformulated in terms of six inequalities \cite{Sauer_2021}:
 \begin{equation}
 \label{eq:CGineq}
0\leqslant 2-\frac{4 \lambda_i^2+\lambda_j^2}{|\lambda_i|},\quad \lambda^2_i \geqslant \lambda^2_j \geqslant \lambda^2_k, \quad i \neq j \neq k,
 \end{equation}
where $i,j,k \in \{1,2,3\}$ and $\lambda_i$ is an eigenvalue of $C_\nu$.  For any quantum state $\rho_{\text{AB}}$
that satisfies this condition, no configuration of the six measurement settings can result in a violation of the CG inequality.

The relationship between the CHSH and CG inequalities is nontrivial. As originally observed by Collins and Gisin \cite{CG}, there exist quantum states that violate one of these inequalities while satisfying the other. Moreover, as shown in \cite{Sauer_2021}, the CG inequality is generally more sensitive to entanglement, allowing a larger set of two-qubit entangled states to violate it compared to the CHSH inequality. However, in the specific case of Bell-diagonal states, the opposite is true: more entangled states violate the CHSH inequality than the Collins–Gisin inequality.

\section{Results}
\label{IV}

In this section, we present the quantitative results for the two game configurations, MPMG and DG, highlighting how entanglement provides measurable advantages over classical strategies. We identify the thresholds at which quantum advantage emerges and compute the volumes of the corresponding regions within the convex set of quantum states. Additionally, the CHSH and CG inequalities define distinct non-empty convex subsets, each one strictly larger than the set of separable states. The quantum states within these subsets do not violate the respective inequality. Accordingly, we define $\mathcal{R} \subseteq \mathcal{E}$ as the set of entangled states that either yield quantum advantage in one of the two games or violate one of the two Bell inequalities.

Since we restrict our analysis to modified Werner and Bell-diagonal states, the measure of the set of entangled states satisfies $V(\mathcal{E}) = V(\mathcal{S})/2$ (see 
Ref.~\cite{Sauer_2021}). This relation can be verified using the PPT criterion, which provides a complete characterization of separable states in the two-qubit case. As a result, from Eqs. \eqref{eq:Pr} and \eqref{eq:PrQ}, it follows that $P_Q(\mathcal{R}) = 2 P(\mathcal{R})$. We begin by presenting the results for the modified Werner state, which offers a transparent relationship between the games and the Bell inequalities, as the space of quantum states is represented by the interval $[0,1]$. This is followed by the more general case of Bell-diagonal states, which include the modified Werner state as a special case. 

\subsection{Modified Werner states}

In the MPMG using the modified Werner state, the average probability of winning is given by
\begin{equation}
    \overline{w}(F)=\frac{40}{81}F^2+ \frac{4}{81} F +\frac{37}{81}.
\end{equation}
A quantum advantage is achieved when $\overline{w}(F)>8/9$. 
We define the threshold fidelity $F^{\text{MPMG}}_{min}$ as the solution to
$\overline{w}(F^{\text{MPMG}}_{min})=8/9$, which yields
\begin{equation}
    F^{\text{MPMG}}_{min} = \frac{3 \sqrt{39}-1}{20} \approx 0.886,
\end{equation}
and $\mathcal{R_{\text{MPMG}}}=(F^{\text{MPMG}}_{min},1]$. Furthermore, we have $\mathcal{S}=[0,1]$ and $\mathcal{E}=(0.5,1]$, which yields for Eq. \eqref{eq:PrQ} that
\begin{equation}
\label{eq:MPMGWerner}
 P_Q(\mathcal{R_{\text{MPMG}}}) \approx 0.226.  
\end{equation}
For the DG using the modified Werner state, the average winning probability is given by
\begin{equation}
 \overline{w}(F)=\frac{1}{45}\left(24F^2+21\right).  
\end{equation}
A quantum advantage is given when $\overline{w}(F)>13/15$, which occurs for fidelities exceeding the threshold
\begin{equation}
   F^{\text{DG}}_{min} = \frac{\sqrt{3}}{2} \approx 0.866. 
\end{equation}
Hence, $\mathcal{R_{\text{DG}}}=(F^{\text{DG}}_{min},1]$ and
\begin{equation}
\label{eq:DGWerner}
 P_Q(\mathcal{R_{\text{DG}}}) \approx 0.268.   
\end{equation}
In the case of Bell inequalities, one must compute the eigenvalues of the matrix $C_\nu$ in Eq. \eqref{eq:Cn}. These eigenvalues are given in Appendix \ref{App:B} and read:
\begin{equation}
\lambda_1 = \frac{2}{3}F - \frac{1}{6}, \quad
\lambda_2 = \frac{1}{6} - \frac{2}{3}F, \quad
\lambda_3 = \frac{2}{3}F - \frac{1}{6}.
\end{equation}
Substituting into the criterion for no CHSH violation
in Eq. \eqref{eq:horodeckicond}, we obtain
\begin{equation}
    \frac{1}{18} (1-4F)^2 \leqslant \frac{1}{4}.
\end{equation}
Solving this inequality yields the threshold fidelity for CHSH violation:
\begin{equation}
   F^{\text{CHSH}}_{min}=\frac{1}{8}(2+ 3\sqrt{2})\approx 0.78.
\end{equation}
Therefore, quantum states with fidelities in the range $\mathcal{R_{\text{CHSH}}}=(F^{\text{CHSH}}_{min},1]$ violate the CHSH inequality for at least one configuration of the four possible measurement settings. Thus,
\begin{equation}
 P_Q(\mathcal{R_{\text{CHSH}}}) \approx 0.439.   
\end{equation}
For the CG inequality, we observe that the eigenvalues of the matrix $C_\nu$ satisfy
\begin{equation}
    \lambda_1=\lambda_3=-\lambda_2,
\end{equation}
and we define $\lambda=\lambda_1$. With this, the expression
$4\lambda_i^2+\lambda_j^2$ evaluates to
\begin{equation}
  4\lambda_i^2+\lambda_j^2=5\lambda^2 \quad \forall i,j \in \{1,2,3\}.
\end{equation}
Substituting into Eq. \eqref{eq:CGineq}, we obtain:
\begin{equation}
2 - 5|\lambda| \geqslant 0.
\end{equation}
Using $\lambda = \frac{2}{3}F - \frac{1}{6}$, this inequality leads to the condition:
\begin{equation}
F \in \left[0,\,\frac{17}{20}\right].
\end{equation}
Thus, the threshold fidelity for violating CG inequalities is
\begin{equation}
F^{\text{CG}}_{min} = \frac{17}{20}.
\end{equation}
Quantum states with fidelities in the range $\mathcal{R_{\text{CG}}}=(F^{\text{CG}}_{min},1]$ violate the CG inequality for at least one configuration among the six possible measurement settings and we have
\begin{equation}
P_Q(\mathcal{R_{\text{CG}}}) = 0.3.
\end{equation}
To summarize, the regions associated with the different games and Bell inequalities satisfy the following inclusion relations:
\begin{equation}
\label{eq:inclusion}
  \mathcal{R_{\text{MPMG}}} \subset \mathcal{R_{\text{DG}}} \subset \mathcal{R_{\text{CG}}} \subset \mathcal{R_{\text{CHSH}}}.
\end{equation}
Their relative sizes, given by the ratios of their measures with respect to the CHSH region, are approximately:
\begin{eqnarray}
    &&\frac{P_Q(\mathcal{R_{\text{MPMG}}})}{P_Q(\mathcal{R_{\text{CHSH}}})}\approx 0.515, \quad \frac{P_Q(\mathcal{R_{\text{DG}}})}{P_Q(\mathcal{R_{\text{CHSH}}})}\approx 0.61, \nonumber \\
    && \frac{P_Q(\mathcal{R_{\text{CG}}})}{P_Q(\mathcal{R_{\text{CHSH}}})}\approx 0.683. \label{eq:gratio}
\end{eqnarray}

\subsection{Bell-diagonal states}

\begin{figure}[htbp]
\centering
\begin{subfigure}[t]{0.5\textwidth}
\centering
\includegraphics[width=\linewidth]{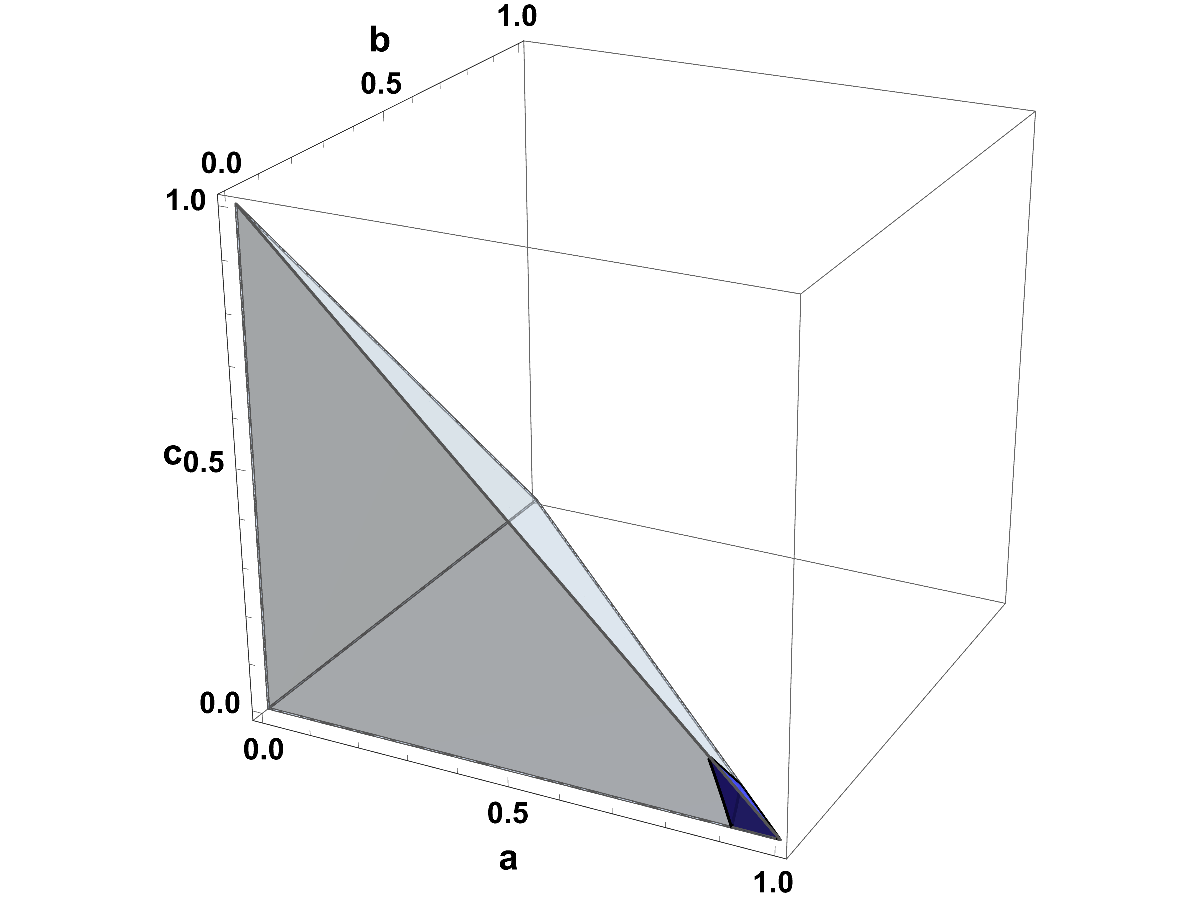}
\caption{$\mathcal{S} \cup \mathcal{R_{\text{MPMG}}}$}
\end{subfigure}
\hfill
\begin{subfigure}[t]{0.5\textwidth}
\centering
\includegraphics[width=\linewidth]{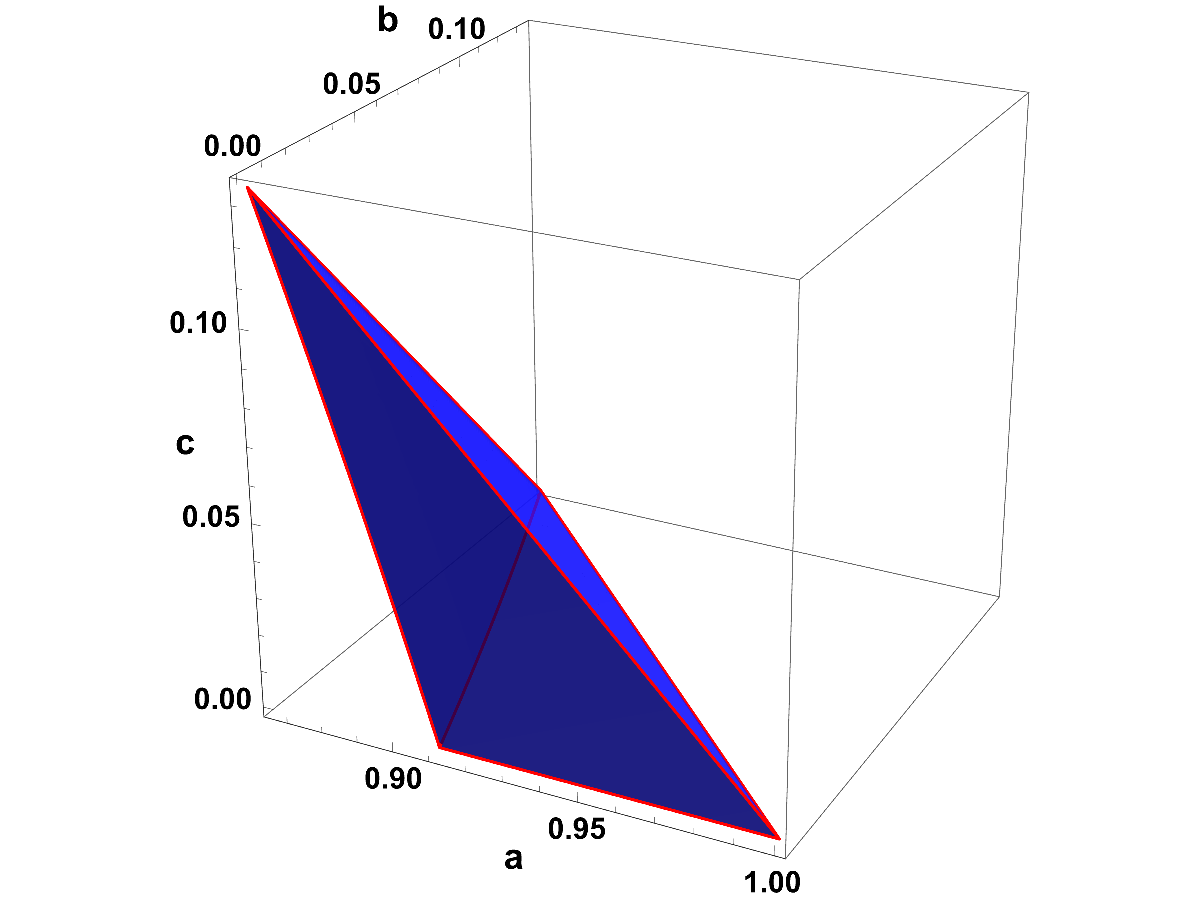}
\caption{$\mathcal{R_{\text{MPMG}}}$}
\end{subfigure}
\caption{Schematic illustrations of (a) $\mathcal{S} \cup \mathcal{R_{\text{MPMG}}}$ and (b) $\mathcal{R_{\text{MPMG}}}$ are shown. The tetrahedron represents the convex set of all Bell-diagonal states, with its four vertices corresponding to the maximally entangled Bell states, see Eq. \eqref{eq:Belldiagonal}. When $a = 1$, the average probability of winning the MPMG is equal to one.}
\label{fig:regions_PM}
\end{figure}

When using a Bell-diagonal state in the MPMG, the average probability of winning is:
\begin{eqnarray}
 \overline{w}(a,b,c) &=& \frac{8}{9}a^{2}
+ \frac{4}{9}(2b+2c-1)\,a
+ \frac{4}{9}b^{2} \nonumber \\
&+& \frac{4}{9}(2c-1)\,b
+ \frac{4}{9}c(c- 1)+ \frac{5}{9}. \label{eq:MPMGwin}  
\end{eqnarray}
The convex set of all Bell-diagonal states, along with the region exhibiting quantum advantage in the MPMG, are defined as follows:
\begin{eqnarray}
 \mathcal{S} &=& \left\{(a,b,c) \in \mathbb{R}^3 \mid 
0 \leqslant a,b,c \leqslant 1, \right. \nonumber \\
&\quad & \left. a+b+c \leqslant 1 \right\},  \\
\mathcal{R_{\text{MPMG}}} &=& \mathcal{S} \cap 
\left\{(a,b,c) \in \mathbb{R}^3 \mid 
\frac{8}{9} < \overline{w}(a,b,c) \leqslant 1 \right\}. \nonumber
\end{eqnarray}
A detailed mathematical characterization of $\mathcal{R_{\text{MPMG}}}$ is provided in Appendix \ref{App:A}. Based on this description, numerical integration yields
\begin{equation}
\label{eq:MPMGBELL}
P_Q(\mathcal{R_{\text{MPMG}}}) \approx 0.003.
\end{equation}
Interestingly, with the growth of the parameter space describing the quantum systems, the ratio in Eq. \eqref{eq:MPMGBELL} becomes significantly smaller than that in Eq. \eqref{eq:MPMGWerner}. This reduction is substantial even considering that entangled states constitute half of the total state space for both the modified Werner and Bell-diagonal states. Fig. \ref{fig:regions_PM} displays the region $\mathcal{R_{\text{MPMG}}}$ within the tetrahedron, which represents the state space of the Bell-diagonal states, and shows that this region is not convex.

\begin{figure}[htbp]
\centering
\begin{subfigure}[t]{0.5\textwidth}
\centering
\includegraphics[width=\linewidth]{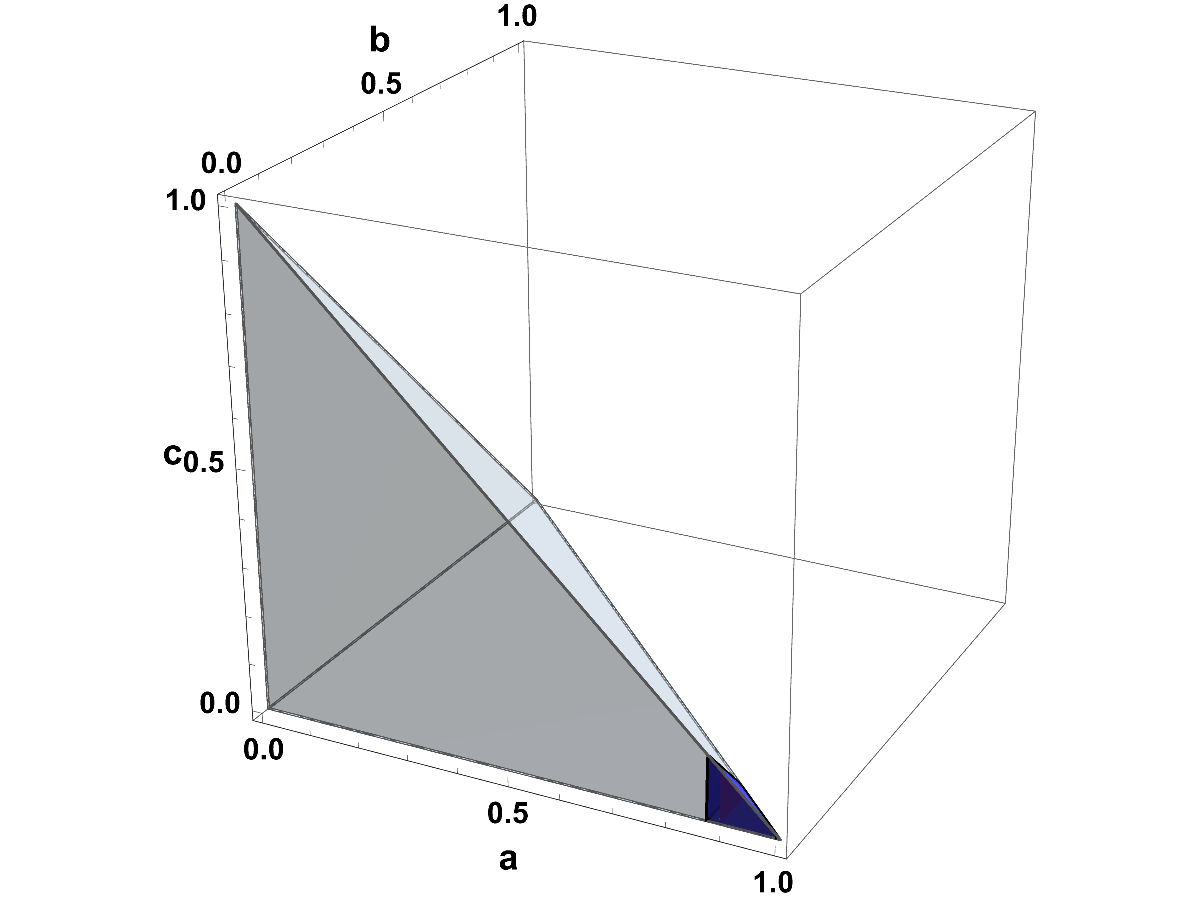}
\caption{$\mathcal{S} \cup \mathcal{R_{\text{DG}}}$}
\end{subfigure}
\hfill
\begin{subfigure}[t]{0.5\textwidth}
\centering
\includegraphics[width=\linewidth]{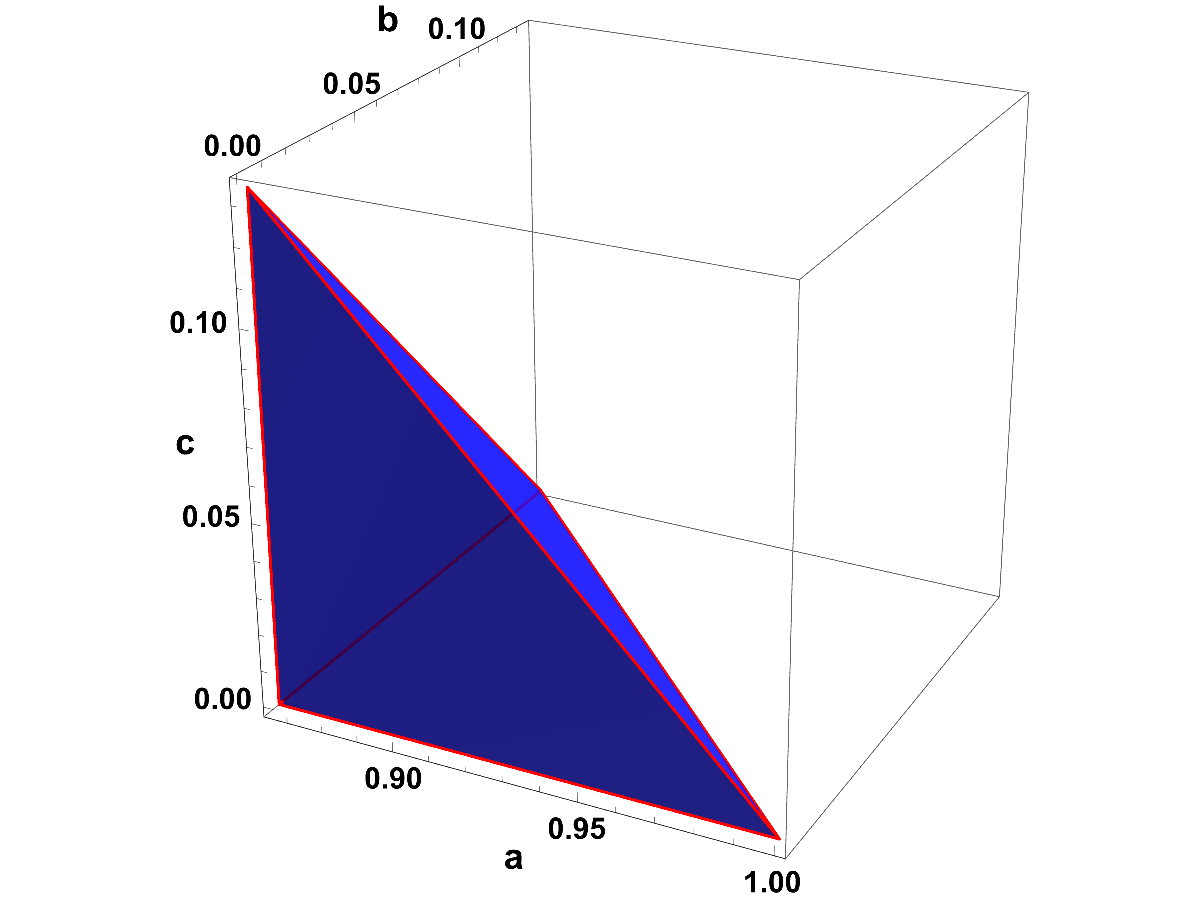}
\caption{$\mathcal{R_{\text{DG}}}$}
\end{subfigure}
\caption{Schematic illustrations of (a) $\mathcal{S} \cup \mathcal{R_{\text{DG}}}$ and (b) $\mathcal{R_{\text{DG}}}$ are shown. For the description of the tetrahedron, see Fig. \ref{fig:regions_PM}. When $a = 1$, the average probability of winning the DG is equal to one.}
\label{fig:regions_D}
\end{figure}

In the case of the DG, the average probability of winning is given by
\begin{equation}
\overline{w}(a,b,c) = \overline{w}(a) = \frac{1}{45} (24 a^2 + 21).
\end{equation}
We immediately observe that this probability depends only on $a$, in contrast to the probability obtained for the MPMG in Eq.~\eqref{eq:MPMGwin}. This arises from the fact that, in the doily game, all three Bell states $\ket{\Psi-}$, $\ket{\Psi+}$, and $\ket{\Phi-}$ contribute equally to the average error probability. Consequently, the single parameter $a$, corresponding to the winning Bell state $\ket{\phi^+}$, is sufficient to determine the average winning probability. This is not the case for the MPMG, where only a selected subset of the possible local operators is used. The region exhibiting quantum advantage is defined as
\begin{equation}
\mathcal{R_{\text{DG}}} = \left\{(a,b,c) \in \mathcal{S} \mid\, \frac{13}{15} < \overline{w}(a) \leq 1 \right\}.
\end{equation}
Our numerical estimation yields
\begin{align}
\label{eq:DGBELL}
P_Q(\mathcal{R_{\text{DG}}}) \approx 0.004.
\end{align}
Since the only constraint is $a > \sqrt{3}/2$, the region $\mathcal{R_{\text{DG}}}$ is convex, and it is illustrated in Fig. \ref{fig:regions_D}. The result in Eq. \eqref{eq:DGBELL}, together with Fig. \ref{fig:regions_D}, shows that the inclusion $\mathcal{R_{\text{MPMG}}} \subset \mathcal{R_{\text{DG}}}$ holds also for Bell-diagonal states. However, the ratio in Eq. \eqref{eq:DGBELL} is also significantly smaller than that in Eq. \eqref{eq:DGWerner}. 

\begin{figure}[htbp]
\centering
\includegraphics[width=0.36\textwidth]{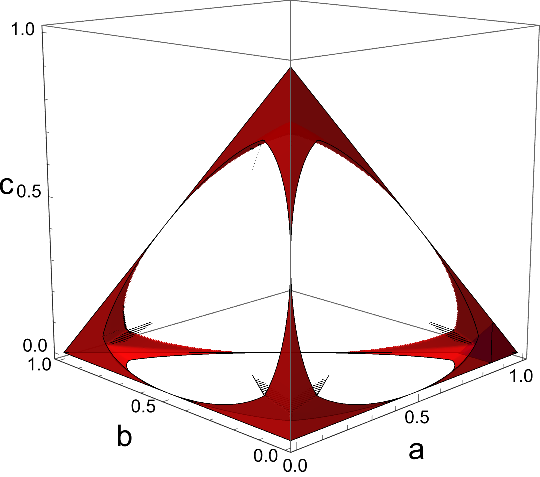}
\caption{Schematic illustrations of the tetrahedron excluding the Steinmetz solid, representing the region $\mathcal{R_{\text{CHSH}}}$ associated with the CHSH inequality. The dark red region at the vertex $a=1$, $\mathcal{R_{\text{DG}}}$, indicates the quantum states that provide a quantum advantage in the DG.  For comparison, see Fig. \ref{fig:regions_D}. (The small gray spikes in the corners are just artifacts of the rendering algorithm.)}
\label{fig:Ronly_CHSH_2}
\end{figure}

\begin{figure}[htbp]
\centering
\includegraphics[width=0.5\textwidth]{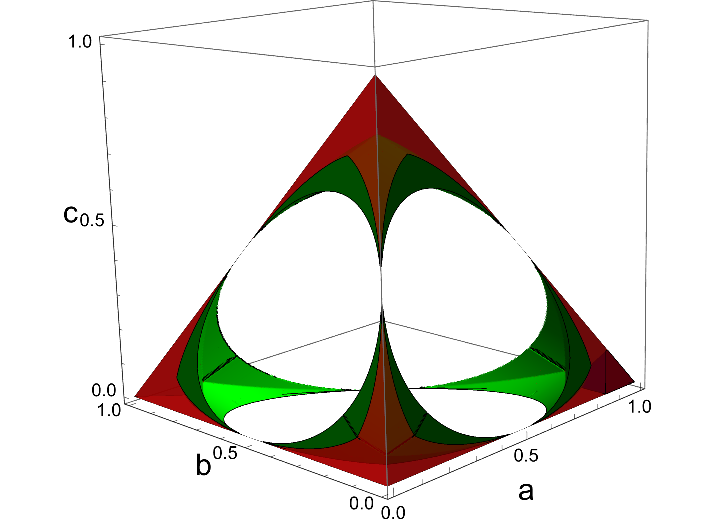}
\caption{Schematic illustrations of the tetrahedron excluding the Steinmetz solid, showing the region $\mathcal{R_{\text{CHSH}}}$ associated with the CHSH inequality, along with the convex body corresponding to the CG inequality. The green region represents the set difference $\mathcal{R_{\text{CHSH}}} \setminus \mathcal{R_{\text{CG}}}$. For comparison, see Fig. \ref{fig:Ronly_CHSH_2}. The black lines mark the boundaries of the six regions defined by Eq. \eqref{eq:CGineq}.}
\label{fig:Ronly_CHSH_CG}
\end{figure}

Finally, we investigate the CHSH and the CG inequalities based on Eqs. \eqref{eq:abcCHSH} and \eqref{eq:CGineq}. Our numerical estimations yield
\begin{equation}
  P_Q(\mathcal{R_{\text{CHSH}}}) \approx 0.175,
\end{equation}
and
\begin{equation}
    P_Q(\mathcal{R_{\text{CG}}}) \approx 0.073.
\end{equation}
We note that both ratios reflect the fact that entangled states violating a Bell inequality are still more typical than those that provide a quantum advantage in the respective games. Since we have already identified that $\mathcal{R_{\text{MPMG}}} \subset \mathcal{R_{\text{DG}}}$, we now illustrate graphically how these regions relate to each other. In Fig. \ref{fig:Ronly_CHSH_2}, one can observe that $\mathcal{R_{\text{CHSH}}}$ is significantly larger than $\mathcal{R_{\text{DG}}}$. The four vertices of the tetrahedron, corresponding to the Bell states, and the associated regions lie outside the Steinmetz solid, meaning they include states capable of violating a CHSH inequality. The quantum advantage in the DG is concentrated around a single vertex, specifically $a = 1$, which corresponds to the $\ket{\Phi^+}$ state. However, even in this corner of the tetrahedron, the set of CHSH violating states remains larger. When Alice and Bob know the actual parameters of the Bell-diagonal state that they possess,, they can use local operations to transform it so that $\ket{\Phi^+}$ has the largest weight in Eq. \eqref{eq:Belldiagonal}. Therefore, the regions close to the other three corners in Fig. \ref{fig:Ronly_CHSH_2} can also be considered as regions where the doily game can detect entanglement. However, even after multiplying the results in Eqs. \eqref{eq:MPMGBELL} and \eqref{eq:DGBELL} by four, the resulting ratios remain significantly smaller than those associated with the two Bell inequalities.

The CG inequality also defines a convex region that contains not only all separable states but also some entangled ones. According to Eq. \eqref{eq:CGineq}, this region can be decomposed into six parts. In Fig. \ref{fig:Ronly_CHSH_CG}, the region $\mathcal{R_{\text{DG}}}$ is shown, including states that can violate a CHSH inequality but do not violate any CG inequality. From the figure, we observe that $\mathcal{R_{\text{DG}}} \subset \mathcal{R_{\text{CG}}} \subset \mathcal{R_{\text{CHSH}}}$. However, as previously mentioned, the relationship between the CHSH and CG inequalities is not straightforward for general two-qubit states. Although the inclusion relations in Eq. \eqref{eq:inclusion} remain valid, the relative volumes of the regions have decreased, as shown in Table \ref{tab_parameters}. For Bell-diagonal states, their relative sizes, measured with respect to the CHSH region, are approximately:
\begin{eqnarray}
    &&\frac{P_Q(\mathcal{R_{\text{MPMG}}})}{P_Q(\mathcal{R_{\text{CHSH}}})}\approx 0.017, \quad \frac{P_Q(\mathcal{R_{\text{DG}}})}{P_Q(\mathcal{R_{\text{CHSH}}})}\approx 0.022, \nonumber \\
    && \frac{P_Q(\mathcal{R_{\text{CG}}})}{P_Q(\mathcal{R_{\text{CHSH}}})}\approx 0.417. \label{eq:gratio2}
\end{eqnarray}

These results, when compared with those for the modified Werner state in Eq. \eqref{eq:gratio}, show that Bell inequalities are less affected by the increase in parameter space than the quantum advantage observed in both games. The typicality of  two-qubit entangled states that provide a quantum advantage decreases significantly more than that of states capable of violating a Bell inequality.

\begin{table}[ht!]
    \centering
    \begin{tabular}{|c|c|c|}
    \hline
       -- & Werner & Bell-diagonal  \\ \hline 
     MPMG   &  0.226  &   0.003 \\ \hline
     DG & 0.268  & 0.004 \\ \hline
      CG & 0.3  & 0.073 \\
   \hline 
   CHSH & 0.439 & 0.175 \\
   \hline
    \end{tabular}
    \caption{Estimates of $P_Q(\mathcal{R})$ from Eq. \eqref{eq:PrQ} are provided for both families of 
    two-qubit quantum states, for both games, and for both Bell inequalities.}
    \label{tab_parameters}
\end{table}

\section{Discussion}
\label{V}

Quantum pseudo-telepathy games offer alternative formulations of Bell’s argument against noncontextual hidden variable models. They are often regarded as more compelling than traditional Bell-type arguments \cite{Brassard2005}, which impose nontrivial constraints on the correlations of spatially separated particles. In this context, we quantitatively investigated the typicality of  two-qubit entangled states that either provide a quantum advantage in a game or can violate a Bell inequality.

As examples, we considered the Mermin–Peres magic square game (MPMG) and the doily game (DG), and compared their performance with that of the Clauser–Horne–Shimony–Holt (CHSH) and Collins–Gisin (CG) inequalities. Our analysis focused on two representative families of two-qubit states: modified Werner states and Bell-diagonal states. For each method and state family, we identified and quantified the regions in the quantum state space where quantum game strategies outperform all classical ones or where the quantum states can violate a Bell inequality. These regions were also expressed as a fraction of the entangled state space. This allowed us to evaluate not only the individual performance of each approach but also their relative effectiveness, based on how their corresponding regions relate to one another.

The results show that in the more complex DG, a larger portion of entangled states can provide a quantum advantage compared to the MPMG. This indicates that the DG is more robust to noise and less sensitive to imperfections in the quantum states. However, both Bell inequalities define significantly larger regions than the two games. For the specific families of states considered, the region corresponding to the CHSH inequality is always larger than that of the CG inequality. This relationship between the two Bell-inequalities, however, changes when the full parametrization of two-qubit states is taken into account, as shown in Ref. \cite{Sauer_2021}.

Furthermore, we have shown that, although the ratio between separable and entangled states remains constant within these families of two-qubit states, the regions corresponding to all four approaches begin to shrink when moving from one-parameter quantum states to three-parameter ones. This effect was previously observed for the two Bell inequalities studied in Ref.~\cite{Sauer_2021}, but we found that the shrinkage from the modified Werner states to Bell-diagonal states is considerably more pronounced in the case of quantum games. In this context, although quantum pseudo-telepathy games may provide a more intuitive argument than Bell inequalities, particularly for non-experts, the fraction of two-qubit entangled states that offer a quantum advantage in these games is smaller than the fraction of entangled states that violate a CHSH or CG inequality. Extending the analysis to other families of two-qubit states, and ultimately to the full fifteen-dimensional parameterization \cite{Bengtsson}, will require a generalization of the approach presented here.

\subsection*{Acknowledgments}

We are indebted to F. Petri, M. Schilling, C.-K. Law, \'A. T\'oth, M. Koniorczyk and B. Robotka for several illuminating conversations and discussions. In addition, the authors would like to thank the anonymous reviewers for their helpful and constructive comments, which greatly contributed to the improvement of this paper. G. H. and A. B. thanks the ”Frontline” Research Excellence Programme of the NKFIH (Grant no. KKP133827). J.Z.B. was supported by the Hungarian National Research, Development and Innovation Office within the Quantum Information
National Laboratory of Hungary grants no. 2022-2.1.1-NL-2022-00004 and 134437.

\appendix
\section{Mathematical definition of the region $\mathcal{R_{\text{MPMG}}}$}
\label{App:A}

In this Appendix, we provide the mathematical definition of the region $\mathcal{R_{\text{MPMG}}}$, which is determined by the quantum advantage in the MPMG. To define the boundaries of this geometric region, we considered the equality $\overline{w}(a,b,c) = 8/9$ from Eq. \eqref{eq:MPMGwin}.

\begin{widetext}
\begin{equation}
\resizebox{.95\textwidth}{!}{$
\mathcal{R_{\text{MPMG}}} \coloneqq
\left\{
\begin{array}{@{}l@{\quad}l@{\quad}l@{}}
\mathbf{(I)} & a = \frac{\sqrt{3}}{2}: &
\begin{cases}
b = 0, & c = 1 - \frac{\sqrt{3}}{2}, \\[0.3em]
0 < b < 1 - \frac{\sqrt{3}}{2}, & c = -b - \frac{\sqrt{3}}{2} + 1, \\[0.3em]
b = 1 - \frac{\sqrt{3}}{2}, & c = 0
\end{cases}
\\[1em]
\mathbf{(II)} & \frac{\sqrt{3}}{2} < a < \frac{1}{4} + \frac{\sqrt{7}}{4}: &
\begin{cases}
b = 0, & c = \frac{1}{2} - a + \sqrt{1 - a^2}, \\[0.3em]
b = 0, & \frac{1}{2} - a + \sqrt{1 - a^2} < c < 1 - a, \\[0.3em]
b = 0, & c = 1 - a, \\[0.3em]
0 < b < \frac{1}{2} - a + \sqrt{1 - a^2}, & c = \frac{1}{2} - a - b + \sqrt{1 - a^2}, \\[0.3em]
0 < b < \frac{1}{2} - a + \sqrt{1 - a^2}, & \frac{1}{2} - a - b + \sqrt{1 - a^2} < c < 1 - a - b, \\[0.3em]
0 < b < \frac{1}{2} - a + \sqrt{1 - a^2}, & c = -a - b + 1, \\[0.3em]
b = \frac{1}{2} - a + \sqrt{1 - a^2}, & c = 0, \\[0.3em]
b = \frac{1}{2} - a + \sqrt{1 - a^2}, & 0 < c < \frac{1}{2} - \sqrt{1 - a^2}, \\[0.3em]
b = \frac{1}{2} - a + \sqrt{1 - a^2}, & c = \frac{1}{2} - \sqrt{1 - a^2}, \\[0.3em]
-a + \frac{1}{2} + \sqrt{1 - a^2} < b < 1 - a, & c = 0
\end{cases}
\\[1em]
\mathbf{(III)} & a = \frac{1}{4} + \frac{\sqrt{7}}{4}: &
\begin{cases}
b = 0, & c = 0, \\
b = 0, & 0 < c \leqslant \frac{3}{4} - \frac{\sqrt{7}}{4}, \\
0 < b \leqslant \frac{3}{4} - \frac{\sqrt{7}}{4}, & c = 0, \\
0 < b < \frac{3}{4} - \frac{\sqrt{7}}{4}, & 0 < c < -b + \frac{3}{4} - \frac{\sqrt{7}}{4}, \\
0 < b < \frac{3}{4} - \frac{\sqrt{7}}{4}, & c = -b + \frac{3}{4} - \frac{\sqrt{7}}{4}
\end{cases}
\\[1em]
\mathbf{(IV)} & \frac{1}{4} + \frac{\sqrt{7}}{4} < a < \frac{\sqrt{55}}{8}: &
\begin{cases}
b = 0, & 0 < c < 1 - a, \\
b = 0, & c = 1 - a, \\
0 < b < 1 - a, & c = 0, \\
0 < b < 1 - a, & 0 < c < -a - b + 1, \\
0 < b < 1 - a, & c = -a - b + 1, \\
b = 1 - a, & c = 0
\end{cases}
\\[1em]
\mathbf{(V)} & a = \frac{\sqrt{55}}{8}: &
\begin{cases}
b = 0, & 0 < c \leqslant 1 - \frac{\sqrt{55}}{8}, \\
0 < b \leqslant 1 - \frac{\sqrt{55}}{8}, & c = 0, \\
0 < b < 1 - \frac{\sqrt{55}}{8}, & 0 < c < -b - \frac{\sqrt{55}}{8} + 1, \\
0 < b < 1 - \frac{\sqrt{55}}{8}, & c = -b - \frac{\sqrt{55}}{8} + 1
\end{cases}
\\[1em]
\mathbf{(VI)} & \frac{\sqrt{55}}{8} < a \leqslant 1: &
\begin{cases}
b = 0, & 0 < c \leqslant 1 - a, \\
b = 0, & c = 1 - a, \\
0 < b \leqslant 1 - a, & c = 0, \\
0 < b \leqslant 1 - a, & 0 < c \leqslant -a - b + 1, \\
0 < b \leqslant 1 - a, & c = -a - b + 1, \\
b = 1 - a, & c = 0
\end{cases}
\end{array}
\right.
$}
\label{eq:R_cases}
\end{equation}
\end{widetext}

\section{Modified Werner state in the form of Eq. \eqref{eq:genrho}}
\label{App:B}

The modified Werner state, expressed in the canonical basis $\{\ket{00},\ket{01},\ket{10},\ket{11}\}$, is given by:
\begin{equation}
\rho(F)=\begin{pmatrix}
\frac{1}{3}F+\frac{1}{6} & 0 & 0 & \frac{2}{3}F-\frac{1}{6} \\
0 & \frac{1}{3}-\frac{1}{3}F & 0 & 0 \\
0 & 0 & \frac{1}{3}-\frac{1}{3}F & 0 \\
\frac{2}{3}F-\frac{1}{6} & 0 & 0 & \frac{1}{3}F+\frac{1}{6}
\end{pmatrix}.
\end{equation}

A Bell-diagonal state written in the form of Eq. \eqref{eq:genrho} reads 
\begin{eqnarray}
&&\rho(a_x,a_y,a_z)= \nonumber \\
&&=\begin{pmatrix}
\frac{1}{4}+\frac{1}{2}a_z & 0 & 0 & \frac{1}{2}a_x-\frac{1}{2}a_y \\
0 & \frac{1}{4}-\frac{1}{2}a_z & \frac{1}{2}a_x+\frac{1}{2}a_y & 0 \\
0 & \frac{1}{2}a_x+\frac{1}{2}a_y & \frac{1}{4}-\frac{1}{2}a_z & 0 \\
\frac{1}{2}a_x-\frac{1}{2}a_y & 0 & 0 & \frac{1}{4}+\frac{1}{2}a_z
\end{pmatrix}, \nonumber 
\end{eqnarray}
where $a_x$, $a_y$, and $a_z$ are the diagonal entries of $C_\nu$ (cf. Eq. \eqref{eq:Cn}), and the parametrization in Eq. \eqref{eq:Belldiagonal} is related to $a_x$, $a_y$, and $a_z$ as follows:
\begin{equation}
a_x = a + b - \frac{1}{2}, \qquad
a_y = b + c - \frac{1}{2}, \qquad
a_z = a + c - \frac{1}{2}.
\end{equation}

The modified Werner state corresponds to
\begin{equation}
a = F, \qquad b = c = \frac{1-F}{3}.
\end{equation}

Substituting these into the above expressions yields
\begin{equation}
a_x = \frac{2}{3}F - \frac{1}{6}, \quad
a_y = \frac{1}{6} - \frac{2}{3}F, \quad
a_z = \frac{2}{3}F - \frac{1}{6}.
\end{equation}

Hence,
\begin{equation}
a_x = -a_y = a_z.
\end{equation}

\bibliography{main}

\end{document}